\documentclass[10pt,aps,pra,twocolumn,superscriptaddress,floatfix,english]{revtex4-1}
\usepackage{amsmath,amsthm,amsfonts,amssymb}
\usepackage{graphicx,array}
\usepackage[usenames,dvipsnames,svgnames,table]{xcolor}
\usepackage{multirow,bigstrut}
\usepackage[colorlinks=true]{hyperref}
\hypersetup{linkcolor=NavyBlue,urlcolor=NavyBlue,citecolor=NavyBlue}

\newcommand{\lp}{\left(}  
\newcommand{\rp}{\right)} 
\newcommand{\lb}{\left[}   
\newcommand{\rb}{\right]}  

\newcommand{\mb}{\mathbf} 				
\newcommand{\cl}{\mathcal} 				
\newcommand{\tsf}[1]{\textsf{#1}} 		
\DeclareMathOperator{\sinc}{sinc}       

\newcommand{\bd}{\mb{d}} 				
\newcommand{\br}{\mb{r}} 				

\newcommand{\ket}[1]{\left|{#1}\right\rangle}

\newcommand{\bra}[1]{\langle{#1}|}
\newcommand{\braket}[2]{\langle{#1}|{#2}\rangle}

\newcommand{\abs}[1]{\left | #1 \right |}   		

\DeclareMathOperator{\tr}{Tr}
\newcommand{\hp}{\mathsf{H}}

\newcommand{\pemin}{P_{\mathrm{e,\min}}}
\newcommand{\pemink}{P_{\mathrm{e,\min}|L}}
\newcommand{\diff}{{\rm d}}

\newcommand{\eq}{equation}
\newcommand{\eqs}{equations}

\begin{document}
\title{Quantum-optimal detection of one-versus-two incoherent optical sources with arbitrary separation}

\author{Xiao-Ming Lu}
\affiliation{Department of Physics, Hangzhou Dianzi University, Hangzhou 310018, China}
\affiliation{Department of Electrical and Computer Engineering, National University of Singapore, 4 Engineering Drive 3, Singapore 117583, Singapore}

\author{Hari Krovi}
\affiliation{Quantum Information Processing Group, Raytheon BBN Technologies, Cambridge, Massachusetts 02138, USA}

\author{Ranjith Nair}
\affiliation{Department of Electrical and Computer Engineering, National University of Singapore, 4 Engineering Drive 3, Singapore 117583, Singapore}

\author{Saikat Guha}
\affiliation{College of Optical Sciences, University of Arizona, 1630 E. University Blvd., Tucson, Arizona 85719, USA}


\author{Jeffrey H.~Shapiro}
\affiliation{Research Laboratory of Electronics, Massachusetts Institute of Technology, Cambridge, Massachusetts 02139, USA}

\date{\today}
\begin{abstract}
We analyze the fundamental quantum limit of the resolution of an optical imaging system from the perspective of the detection problem of  deciding whether the optical field in the image plane is generated by one incoherent on-axis source with brightness $\epsilon$ or by two $\epsilon/2$-brightness incoherent sources that are symmetrically disposed about the optical axis. 
Using the exact thermal-state model of the field, we derive the quantum Chernoff bound for the detection problem, which specifies the optimum rate of decay of the error probability with increasing number of collected photons that is allowed by quantum mechanics. We then show that  recently proposed linear-optic schemes approach the quantum Chernoff bound---the method of binary spatial-mode demultiplexing (B-SPADE) is quantum-optimal for all values of  separation, while a method using image-inversion interferometry (SLIVER) is  near-optimal for sub-Rayleigh separations. We then simplify our model using a low-brightness approximation that is very accurate for optical microscopy and astronomy, derive quantum Chernoff bounds conditional on the number of photons detected, and show the optimality of our schemes in this conditional detection paradigm. For comparison, we analytically demonstrate the superior scaling of the Chernoff bound for our schemes with source separation relative to that of spatially-resolved direct imaging. Our schemes have the advantages over the quantum-optimal (Helstrom) measurement in that they do not involve joint measurements over multiple modes, and that they do not require the angular separation for the two-source hypothesis to be given \emph{a priori} and  can offer that information as a bonus in the event of a successful  detection. 
\end{abstract}

\maketitle

The influential Rayleigh criterion for imaging resolution~\cite{LordRayleigh1879}, which specifies a minimum
separation for two incoherent light sources to be  distinguishable by a given imaging system, is
based on heuristic notions.
As pointed out by Feynman~\cite[{}][{Sec.~30--4}]{Feynman1963}: ``Rayleigh's criterion is a rough idea in the first place ...'' and 
a better resolution can be achieved ``... if sufficiently careful measurements of the exact intensity distribution over the diffracted image spot can be made ...''
The fundamental measurement noise is the quantum noise necessarily accompanying any measurement.
A more rigorous approach to the resolution measure that accounts for the quantum noise in ideal spatially-resolved image-plane photon counting can be formulated using the classical Cram\'er-Rao bound on the minimum estimation error for locating the sources~\cite{Ram2006,Chao2016}.  
Very recently, using methods of quantum estimation theory~\cite{Helstrom1976,Holevo1982}, it was found that the estimation of the separation between two incoherent sources below the Rayleigh criterion can be drastically  improved by measurements employing pre-detection linear optic processing of the collected light, followed by photon counting~\cite{Tsang2016b,Nair2016,Tsang2016a,Nair2016a,Ang2016,Lupo2016,Rehacek2016,Tsang2016c,KGA17,YNT+17,Tang2016,Yang2016,Tham2017,Paur2016}.

Besides the minimum error of estimating the separation of two point sources, the resolving power of an imaging system can also be studied via the paradigmatic detection problem of deciding whether the optical field in the image
plane is generated by one source or two sources~\cite{Harris1964,Helstrom1973,Acuna1997,Shahram2006,Dutton2010}.
This detection perspective is especially relevant to the  detection of
binary stars and exoplanets~\cite{Acuna1997,Labeyrie2006} and the detection of
protein multimers with fluorescent microscopes~\cite{Nan2013}. 
In a pioneering work~\cite{Helstrom1973}, Helstrom obtained the mathematical description of the
quantum-optimal measurement that minimizes the error probability for detecting one or two point sources emitting quasi-monochromatic thermal light.  Unfortunately, in addition to having no known physical realization, his method
requires the separation between the two hypothetical sources to be
given, though this separation is usually unknown in practice.

Here we investigate the performance of two practical quantum
measurements for the detection of  weak incoherent quasi-monochromatic point light sources.  We assess the performance of these measurements vis-a-vis direct imaging and the optimum quantum measurement using the
asymptotic error exponent (or Chernoff exponent), which specifies the rate at which the error probability decreases exponentially as
the observation time or number of received photons increases.  We show that a binary spatial-mode
demultiplexing (B-SPADE) scheme~\cite{Tsang2016b} is quantum-optimal
for all values of separations in the following two senses: (1) the
asymptotic error exponent attains the maximum allowed by quantum mechanics, and (2) the
error probability of a simple decision rule based on the observations
of the B-SPADE is close to the quantum limit.  We also show that
the scheme of superlocalization by image inversion interferometry
(SLIVER)~\cite{Nair2016,Nair2016a} is near-optimal for sub-Rayleigh separations. 
The Chernoff exponents of both schemes are shown to be superior to that of ideal shot-noise-limited continuum direct imaging in the sub-Rayleigh regime. In addition to
their superiority over direct imaging, our methods do not require the capability to perform joint quantum measurements, do not
require the two-source separation to be known \emph{a priori}, can offer an accurate estimate of
this separation in the event of a successful detection~\cite{Tsang2016b,Nair2016,Tsang2016a,Nair2016a,Ang2016,Lupo2016,Rehacek2016,Tsang2016c},
and rely on methods that have been experimentally demonstrated in the context of parameter
estimation~\cite{Tang2016,Yang2016,Tham2017,Paur2016}. These
 advantages over the Helstrom measurement~\cite{Helstrom1973} hold
tremendous promise for practical detection applications in both
astronomy~\cite{Acuna1997} and molecular imaging~\cite{Nan2013}.
\\

\begin{figure}[tb]
	\centering
	\includegraphics[width=.9\linewidth]{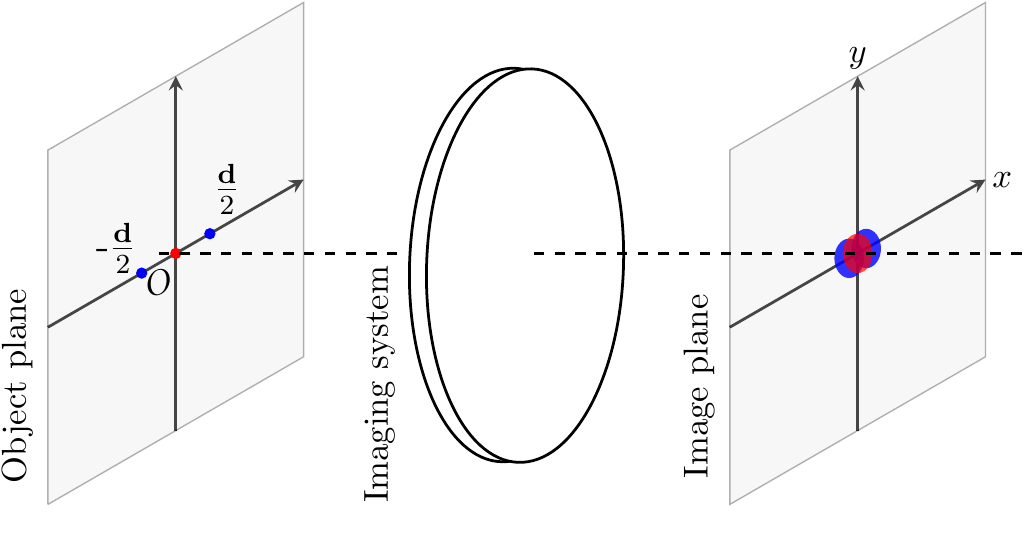}\\
	\caption{Schematic of the imaging of incoherent point light sources by a spatially-invariant imaging system. 
	The images of two point sources of equal brightness (blue) spaced closer than the width of the point-spread function (PSF) of the imaging system are difficult to distinguish from that of one point source of the same total brightness (red) located midway between them.
	}
	\label{fig:model}
\end{figure}

\noindent\textbf{\large Results}\\
{\bf One source versus two sources}.
The set-up considered in this work is illustrated in Fig.~\ref{fig:model}.
Under hypothesis $\tsf{H}_1$, we have a single thermal source of brightness (average photon number per temporal mode) $\epsilon$ imaged at the origin of the image plane. Under  hypothesis $\tsf{H}_2$, we have two thermal sources, each of strength $\epsilon/2$, located a distance $d$ apart and imaged at the points $\pm\bd/2 =(\pm d/2, 0)$ in the image plane. 
To focus on the resolution power of an optical imaging system, the total brightness is assumed to be identical under the two hypotheses so that simple photon counting is ineffective as a decision strategy. Similarly, the sources are presumed to have identical frequency spectra so that spectroscopy cannot help to distinguish the hypotheses.
A strategy for accepting one or the other hypothesis, known as a decision rule, is
given by partitioning the space of observations $\mathcal{Z}$ (which is determined by the choice of measurement) into two disjoint regions
 $\mathcal{Z}_1$ and $\mathcal{Z}_2$; the one-source hypothesis
$\hp_1$ is accepted if the observation belongs to $\mathcal{Z}_1$, and 
 $\hp_2$ is accepted otherwise.  The performed
quantum measurement can be described by a positive-operator-valued
measure (POVM) $\{E(z)\}_{z \in \cl{Z}}$, where $z$ denotes the outcome, and the $E(z)$'s
are nonnegative operators resolving the identity operator as
$\int\!{\rm d}\mu(z) E(z)=\openone$ with $\mu(z)$ being an appropriate
measure on $\mathcal{Z}$~\cite{Helstrom1976,Holevo1982,Hayashi2006}.  
Define $E_1=\int_{z\in\mathcal{Z}_1}\!\diff\mu(z) E(z)$ and $E_2=\int_{z\in\mathcal{Z}_2}\!\diff\mu(z) E(z)$.
Let $\rho_1$ and $\rho_2$ be the density operators for the fields arriving at the image plane per temporal mode under $\hp_1$ and $\hp_2$, respectively. 
Assuming a flat emission spectrum over the bandwidth $W$, the probabilities of
the false-alarm (accepting $\hp_2$ when $\hp_1$ is true) and miss  (accepting $\hp_1$ when $\hp_2$ is true) errors for 
one-source-versus-two testing are given by
$\alpha \equiv \tr(E_2\rho_1^{\otimes M})$
and
$\beta \equiv \tr(E_1\rho_2^{\otimes M})$
respectively, where $M\simeq WT$ with $T$ being the observation time is the number of available temporal modes (also called the \emph{sample size}).
Assuming prior probabilities $p_1$ and $p_2$ for the respective hypotheses, the average probability of error is
\begin{equation}
	P_\mathrm{e} \equiv p_1 \alpha + p_2 \beta,	
\end{equation} 
which is widely used to assess the performance of a quantum decision strategy constituted by a quantum measurement and a classical decision rule~\cite{Helstrom1976}.
The minimum error probability optimized over all quantum decision strategies is given by the Helstrom formula~\cite{Helstrom1976}
\begin{equation}\label{eq:pe_min}
	\pemin =\frac{1}{2}\big(1-\Vert p_2\rho_2^{\otimes M} - p_1\rho_1^{\otimes M}\Vert_1\big),
\end{equation}
where $\Vert A \Vert_{1} \equiv \tr\sqrt{A^\dagger A}$ is the trace norm. 
The minimum error probability can be achieved by the Helstrom-Holevo test in which $E_2$ is taken to be the projector onto the eigen subspace of $p_2\rho_2^{\otimes M} - p_1\rho_1^{\otimes M}$ with positive eigenvalues~\cite{Helstrom1976,Holevo1973b}. We refer to this optimal measurement as the \emph{Helstrom measurement} henceforth.

While the Helstrom formula Eq.~\eqref{eq:pe_min} allows exact computation of the optimum error probability in principle, it is difficult to physically implement  the Helstrom measurement for several reasons.
Firstly, the optimal measurement is a joint one over multiple samples~\cite{Hayashi2006}.
Secondly, this measurement depends on the separation between the two hypothetic point sources, which is often unknown in the first place.
Lastly, the optimal measurement in general depends on the ratio of the prior probabilities of the two hypotheses, whose determination is often subjective.
To circumvent these difficulties, we study the performance of two realizable measurements: B-SPADE~\cite{Tsang2016b} and
SLIVER~\cite{Nair2016}, originally introduced in the context of estimating the separation between two closely-spaced incoherent point sources. \\

\noindent{\bf B-SPADE.}
Spatial-mode demultiplexing refers to spatially separating the image-plane optical field into its components in any chosen set of orthogonal spatial modes~\cite{Tsang2016b}. 
The binary version of spatial-mode demultiplexing, B-SPADE, uses a device that separates a specific spatial mode from all other modes orthogonal to it, and on-off detectors (that can only distinguish between zero and one or more photons) are placed at the two output ports.
In our set-up, the selected spatial mode is chosen to be that generated by the point source at the origin of the object plane.
Such a separation of modes is always possible in principle for any given point-spread function (PSF), and various linear-optics schemes can be envisaged to realize it~\cite{RZB+94,MNJ+10,Mil13}.
\\

\noindent{\bf SLIVER.}
The second practical measurement we consider is SLIVER, which separates the optical field at the image plane into its symmetric and antisymmetric components with respect to  inversion at the origin, followed by on-off photon detection in the respective ports~\cite{Nair2016}. 
Here, we assume that the PSF is reflection-symmetric in the $y$-axis, i.e., $\psi(-x,y) = \psi(x,y)$, and consider a modified SLIVER for which the inversion operation is replaced by the reflection operation about the $y$-axis---this modification corresponds to the Pix-SLIVER scheme of~\cite{Nair2016a} with single-pixel (bucket) on-off detectors at the two outputs. For simplicity, we  refer to this modified version as SLIVER henceforth.
All photo-detectors in both B-SPADE and SLIVER are assumed free from dark counts, or at least that the dark-count rate is so far below the signal-count rate as to be negligible.
\\

\noindent{\bf Asymptotic error (Chernoff) exponents.}
In realistic imaging situations, we usually deal with a large sample size $M \gg 1$, which motivates using the asymptotic error exponent as a useful metric for comparing the performance of different measurement schemes against the Helstrom measurement.
For any specific quantum measurement performed on each sample, it is known that the minimum error probability $\pemin^{\rm (meas)}$ over all decision rules decreases exponentially in $M$ as $\pemin^{\rm (meas)} \sim \exp(-M\xi^{\rm (meas)})$. 
The asymptotic error exponent $\xi^{\rm (meas)} = - \lim_{M \rightarrow \infty}\frac{1}{M} \log \pemin^{\rm (meas)}$ can be given by the Chernoff exponent (also known as Chernoff information or Chernoff distance)~\cite{chernoff1952,VanTrees2013,Cover2012}, namely,
\begin{align}\label{eq:classical_Chernoff}
	\xi^{\rm (meas)} = -\log\min_{0\leq s\leq1} \int\!\mathrm{d} \mu(z)\,\Lambda_1(z)^s \Lambda_2(z)^{1-s}, 
\end{align}
where $\Lambda_j(z)=\tr[E(z)\rho_j]$ is the probability of obtaining the outcome $z$ under the hypothesis $\hp_j$, and $\{E(z)\}$ is the POVM for the measurement.
On the other hand, the error probability $\pemin$ of the optimum quantum measurement  (which is in general a joint measurement on the $M$ samples) scales with the exponent $\xi$ known as  the \emph{quantum Chernoff exponent}, which is given by~\cite{Ogawa2004,Kargin2005,Audenaert2007,Nussbaum2009,Audenaert2008}: 
\begin{align}\label{eq:chernoff}
	 \xi \equiv -\log\min_{0\leq s\leq1}\tr(\rho_1^s\rho_2^{1-s}).
\end{align}
Note that $\xi$ is independent of the measurement and $\xi\geq\xi^{\rm (meas)}$ holds for any measurement.

To calculate the Chernoff exponent, we need to know the characteristics of the imaging system.
Without essential loss of generality, we suppose that the imaging system is spatially-invariant and of unit magnification~\cite{Goo05Fourier} and is described by its 2-D amplitude PSF $\psi(\br)$, where $\br = (x,y)$ is the transverse coordinate in the image plane $\cl{I}$.
Moreover, we take the PSF to be normalized, i.e., $\int_{\cl{I}}\mathrm{d}x\mathrm{d}y\,|\psi(x,y)|^2=1$.
For thermal sources, we show that the exact Chernoff exponents $\xi^{\rm (B-SPADE)}$ and $\xi^{\rm (SLIVER)}$ for  B-SPADE and SLIVER respectively and the quantum Chernoff exponent $\xi$ are  given by (see the Methods)
\begin{align}
	\xi^{\rm (B-SPADE)} &= \xi = \log[(1+\epsilon_-)(1+\epsilon_+ - \mu^2\epsilon_+)], \label{eq:ce-limit-bspade}\\
	\xi^{\rm (SLIVER)}  &= \log(1+\epsilon_-)\leq \xi, \label{eq:ce-sliver}
\end{align}
where 
the $d$-dependent quantities
\begin{align} \label{eq:lambda_mu}
	\epsilon_\pm = \frac{1 \pm \delta(d)}{2}\epsilon 
	\quad\mbox{and}\quad
	\mu = \delta\lp\frac{d}{2}\rp\sqrt{\frac{2}{1+\delta(d)}}
\end{align}
are defined in terms of the overlap function of the PSF for displacements along the $x$ axis:
\begin{align}\label{eq:delta}
	\delta(d) := \int_\mathcal{I} \diff x\diff y \;\psi^*(x,y)\, \psi(x-d,y).
\end{align}
Moreover, we here assume that the overlap function (and hence $\epsilon_\pm$ and $\mu$) is real-valued.
This assumption is satisfied for inversion-symmetric PSFs, i.e., $\psi(x,y)=\psi(-x,-y)$, and $y$-axis reflection-symmetric PSFs, i.e., $\psi(x,y)=\psi(-x,y)$~\cite{Nair2016a,Ang2016}.

It can be seen from \eq~\eqref{eq:ce-limit-bspade} that the Chernoff exponent of the B-SPADE is always equal to the quantum Chernoff exponent, meaning that B-SPADE is asymptotically optimal.
For SLIVER, the Chernoff exponent is in general not quantum-optimal but is close to quantum-optimal in the sub-Rayleigh regime of small $d$, where $\mu$ is close to unity.

We consider three typical kinds of PSFs, corresponding to Gaussian apertures, rectangular hard apertures, and circular hard apertures.
The PSFs can be respectively written as
\begin{align}\label{eq:psf}
\begin{split}
	\psi_{\rm gaus}(x,y) &= \frac{1}{\sqrt{2\pi}\sigma} \exp\left(-\frac{x^2 + y^2}{4\sigma^2}\right), \\
	\psi_{\rm rect}(x,y) &= \frac{1}{\pi\sqrt{\sigma_x\sigma_y}} \mathrm{sinc}(x/\sigma_x)\,\mathrm{sinc}(y/\sigma_y),\mbox{ and}\\
	\psi_{\rm circ}(x,y) &= \frac{1}{2\sqrt{\pi}\sigma_c} \mathrm{jinc}(\sqrt{x^{2}+y^{2}}/\sigma_c),	
\end{split}
\end{align}
where $\sinc(x) \equiv\sin(x)/x$, $\mathrm{jinc}(x) \equiv 2J_{1}(x)/x$, and $J_1(x)$ is the Bessel function of the first kind.
The ``characteristic lengths'' $\sigma$, $\sigma_x$, $\sigma_y$, and $\sigma_c$ are related to the features of apertures as follows.
For a Gaussian aperture, which is commonly assumed in fluorescence microscopy~\cite{Chao2016},  we have $\sigma=\lambda/2\pi\mathrm{NA}$ with $\lambda$ being the free-space center wavelength and NA the effective numerical aperture of the system.
For a $D_x\times D_y$ rectangular aperture, the characteristic length along the $x$ and $y$ directions are given by $\sigma_x=\lambda F / \pi D_x$ and $\sigma_y=\lambda F / \pi D_y$, respectively, where $F$ is the distance between the aperture plane and the image plane in a unit-magnification system. 
For a $D$-diameter circular hard aperture, we have $\sigma_c=\lambda F/\pi D$.
After some algebra, the overlap functions can be shown to be
\begin{align}\label{eq:overlap}
\begin{split}
	\delta_{\rm gaus}(d) &= \exp[-d^2/(8\sigma^2)], \\
	\delta_{\rm rect}(d) &= {\rm sinc}(d/\sigma_x),\mbox{ and }\\
	\delta_{\rm circ}(d) &= {\rm jinc}(d/\sigma_c),
\end{split}
\end{align}
using which the Chernoff exponents can be readily obtained.

\begin{figure}[tb]
	\centering
	\includegraphics[width=1\linewidth]{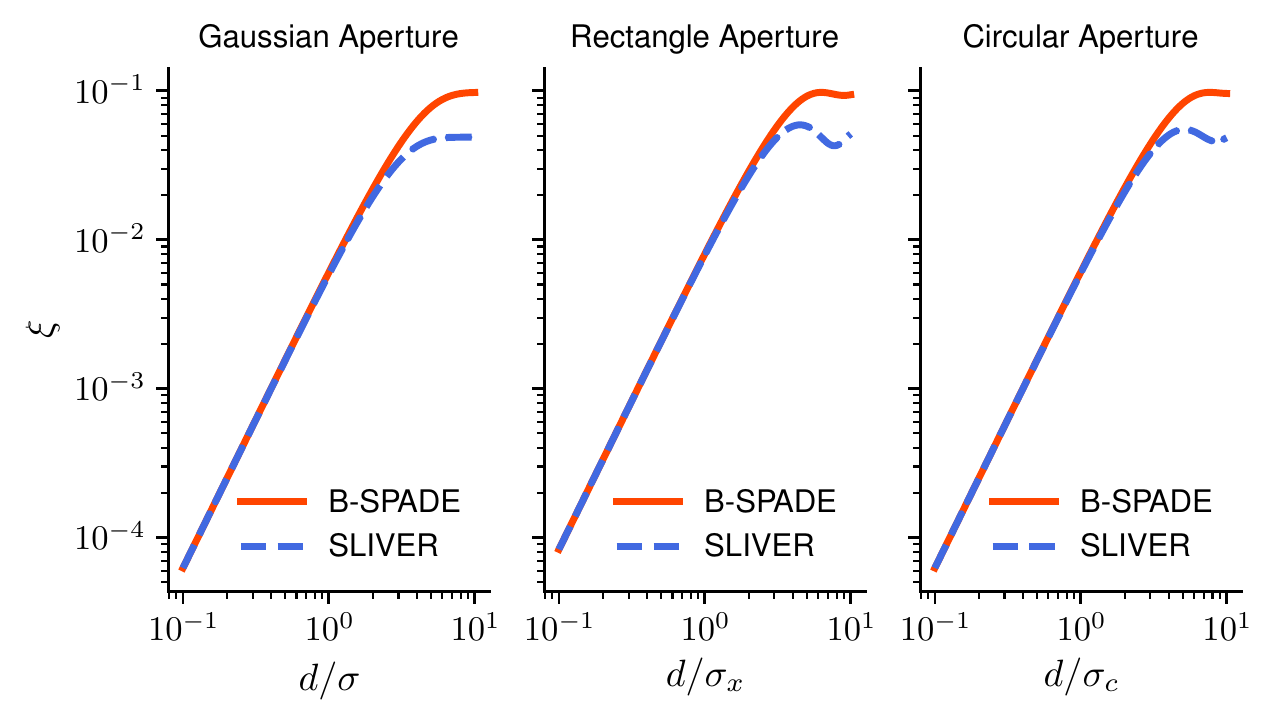}
	\caption{Chernoff exponents for the B-SPADE and SLIVER measurements as a function of the separation $d$. The quantum Chernoff exponent always equals the Chernoff exponent for the B-SPADE measurement. Here, the total strength $\epsilon$ of the thermal sources is set to be $0.1$.}
	\label{fig:chernoff_thermal}
\end{figure}

We plot in Fig.~\ref{fig:chernoff_thermal} the Chernoff exponents of  B-SPADE and SLIVER for the above three PSFs in \eq~(\ref{eq:psf}). 
We can see that in the sub-Rayleigh regime the Chernoff exponents are  insensitive to which PSF is used.
\\

\noindent
{\bf Weak-source model.}
To compare the performance of  B-SPADE and SLIVER with that of direct imaging, we introduce the weak-source model.
In most applications in optical microscopy and astronomy, the source brightness $\epsilon \ll 1$ photons per temporal mode~\cite{Goo85Statistical,Tsang2016b,Gottesman2012,Tsang2011}. 
Then, $\rho_1$ and $\rho_2$ may be considered with high accuracy to be confined to the subspace consisting of zero or one photons, i.e.,
\begin{equation}\label{eq:rhoi01}
	\rho_i \simeq (1-\epsilon)\ket{\rm vac} \bra{\rm vac} + \epsilon \eta_i
\end{equation}
for $i=1,2$, where $\ket{\rm vac}$ denotes the vacuum state and
$\eta_{i}$ are the corresponding one-photon states obtained by neglecting
$O(\epsilon^2)$ terms.  
This approximation enables us to simplify the theory in comparison with Ref.~\cite{Helstrom1973} and still obtain similar results.
The one-photon state for two hypothetical sources can be expressed as
\begin{align}
	\eta(d) &= \frac{\ket{\Psi( d/2)} \bra{\Psi( d/2)}}{2} 
			 + \frac{\ket{\Psi(-d/2)} \bra{\Psi(-d/2)}}{2},
	\nonumber\\
	\ket{\Psi(x')} &\equiv \int_\mathcal{I} \diff x \diff y\,\psi(x- x',y) \ket{x,y}, \label{eq:eta}
\end{align}
where we have introduced the one-photon Dirac kets $\ket{x,y}$ satisfying $\braket{x,y}{x',y'} = \delta(x-x')\delta(y-y')$ and $\int_{\cl{I}} \diff x\diff y \ket{x,y}\bra{x,y} = \openone_1$, the identity on the one-photon subspace in the image plane field~\cite{Tsang2016b}. 
We then have $\eta_1 = \eta(0)$ and $\eta_2=\eta(d)$.

For the specific cases of rectangular and circular hard apertures, Helstrom has derived expressions for the minimum error probability  for thermal light sources~\cite{Helstrom1973}. 
However, these expressions are very complex.
Our weak-source model allows us to simplify the minimum error probability as
\begin{align}
	\pemin &= \sum_{L=0}^{M} \binom{M}{L}(1-\epsilon)^{M-L}\epsilon^L \pemink, 
	\label{eq:binomial}\\
	\pemink &\equiv \frac12-\frac12\Vert p_2\eta_2^{\otimes L} - p_1\eta_1^{\otimes L}\Vert_1.
	\label{eq:pe_conditioned}
\end{align}
Here, $\binom{M}{L}(1-\epsilon)^{M-L}\epsilon^L$ is the
probability of $L$ photons arriving at the imaging plane and $\pemink$
is the minimum probability of error conditioned on detecting  $L$
photons in the image plane.  The form of \eq~\eqref{eq:binomial}
is due to the fact that the distinguishability between $\rho_1$ and
$\rho_2$ lies in the one-photon sector and the zero-photon
event is uninformative.  
It is implicitly assumed in \eq~\eqref{eq:binomial} that the  source flux is low enough that the on-off detectors' recovery time is short compared to the average interarrival time of the photons.
Either the conditional error probability of \eq~\eqref{eq:pe_conditioned} or
the unconditional one of \eq~\eqref{eq:binomial} can be used as a figure of merit, depending on
whether or not the number of the photons arriving at the image plane is measured. 
Helstrom in Ref.~\cite{Helstrom1973} took the latter approach, and the performance was studied with respect to the average total number of photons $N = M \epsilon$ detected over the observation interval. On the other hand, in fluorescence microscopy it is common practice to compare the performance of imaging schemes for the same number of detected photons $L$~\cite{Ram2006,Chao2016}.

We can define a conditional Chernoff exponent $\xi^{\rm (meas)}_{\rm c}$ satisfying 
$\pemink^{\rm (meas)}\sim\exp(-L\xi^{\rm (meas)}_{\rm c})$, which is given by \eq~\eqref{eq:classical_Chernoff} with $\Lambda_j(z)$ replaced by the probability of measurement outcomes conditioned on a photon being detected, i.e.,  $\Lambda_j(z) = \tr[E(z)\eta_j]$.
Similarly, the optimum conditional error probability $\pemink$ decays exponentially with $L$ multiplied by the conditional quantum Chernoff exponent given by
$\xi_{\rm c} \equiv -\log\min_{0\leq s\leq1}\tr(\eta_1^s\eta_2^{1-s})$.
It follows from \eq~\eqref{eq:binomial} that the (unconditional) Chernoff exponents can be obtained via the relation $e^{-\xi} = 1 - \epsilon + \epsilon e^{-\xi_{\rm c}}$ and $e^{-\xi^{\rm (meas)}} = 1 - \epsilon + \epsilon e^{-\xi_{\rm c}^{\rm (meas)}}$.
This implies that the (unconditional) Chernoff exponent is monotonically increasing with the conditional one.
Particularly, we have $\xi^{\rm meas} \simeq \epsilon \xi_{\rm c}^{\rm meas}$ when $\xi^{\rm meas} \ll 1$.
Therefore, we can use either $\xi^{\rm meas}$ or $\xi_{\rm c}^{\rm meas}$ to compare the performance of quantum measurements.

The conditional Chernoff exponents are readily calculated in the weak-source model using \eqs~\eqref{eq:eta}:
\begin{align} 
\xi_{\rm c}^{\rm (B-SPADE)} &= \xi_{\rm c} =  -2 \log \delta(d/2) , \label{eq:cce-bspade}\\
\xi_{\rm c}^{\rm (SLIVER)}  &= - \log \frac{1+\delta(d)}{2} \label{eq:cce-sliver}.
\end{align}
Note that as $d$ decreases, the SLIVER result converges to the B-SPADE result, which can also be seen in Fig.~\ref{fig:chernoff_thermal}.
\\

\noindent{\bf Direct Imaging.}
Direct imaging (DI) using a charge-coupled device (CCD) camera is a standard detection technique in microscopy and telescopy~\cite{Chao2016}. To compare our schemes to direct imaging, we make the conservative assumption of an ideal noiseless photodetector with infinite spatial resolution and unity quantum efficiency placed in the image plane. In the weak-source model, and conditional on a photon being detected in a given temporal mode, the observation consists of its  position of arrival $(x,y) \in \mathcal{I}$. Using \eqs~\eqref{eq:eta}, the resulting probability densities for the observation are $\Lambda_1(x,y)=\Upsilon(x,y;0)$ and $\Lambda_2(x,y)=\Upsilon(x,y;d)$ under $\textsf{H}_1$ and $ \textsf{H}_2$, respectively, where
\begin{align} \label{eq:di}
\Upsilon(x,y;d) &\equiv \frac{\left|\psi(x-d/2,y)\right|^2}{2} 
					 + \frac{\left|\psi(x+d/2,y)\right|^2}{2}.
\end{align}
We show in the Methods that the conditional Chernoff exponent for ideal DI in the weak source model scales as $d^4$ in the interesting  regime of small $d$:
\begin{align} \label{eq:ce-di}
	\xi^{\rm (DI)}_{\rm c} \simeq \frac{d^4}{32}\int_{\cl{I}'} \diff x\diff y \,
	\frac{1}{\Upsilon(x,y;0)} \left[\Upsilon^{(2)}(x,y;0)\right]^2,
\end{align}
where $\Upsilon^{(n)}(x,y;d)$ denotes the $n$-th order partial derivative of $\Upsilon(x,y;d)$ with respect to $d$, and $\mathcal{I}'\equiv\{(x,y)\mid \Upsilon(x,y;0)>0\}$.
In contrast, the conditional Chernoff exponents of B-SPADE and SLIVER in the weak-source model are of order $d^2$, which can be seen by using \eq~\eqref{eq:cce-bspade}--\eqref{eq:cce-sliver} and $\partial \delta(d)/\partial d|_{d=0}=0$.

\begin{figure}[tb]
	\centering
	\includegraphics[]{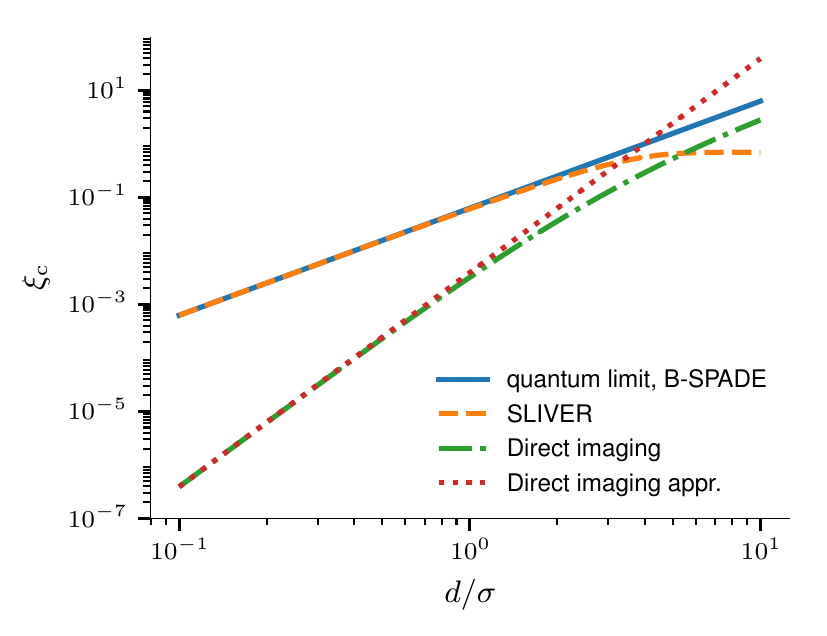}
	\caption{\label{fig:chernoff} 
	Conditional Chernoff exponents in the weak-source model as functions of the normalized two-source separation $d$ for the Gaussian PSF of \eq~\eqref{eq:psf}.
	The Chernoff exponent for  B-SPADE (solid) achieves the quantum limit for all values of $d$, while the Chernoff exponent for  SLIVER (dashed) is near-quantum-optimal for sub-Rayleigh separations.
	In this regime, both schemes outperform  direct imaging --- whose Chernoff exponent is calculated numerically (dash-dotted) as well as by using the small-separation approximation (dotted) given by \eq~\eqref{eq:ce-di} --- by more than an order of magnitude.
	}
\end{figure}

\begin{table}[tb]
    \caption{\label{tab}Conditional Chernoff exponents for the Gaussian PSF.}
	\begin{ruledtabular}
		\begin{tabular}{ll}
        Scheme 			& Chernoff exponent \tabularnewline
        \hline
        B-SPADE/Quantum limit 		& $d^2/16$ \bigstrut \tabularnewline
		SLIVER 			& $d^2/16 - d^4/512 + O(d^6)$ \bigstrut \tabularnewline
        Direct imaging 	& $d^4/256+O(d^6)$ \bigstrut \tabularnewline
        \end{tabular}
	\end{ruledtabular}
\end{table}

The conditional Chernoff exponents for different measurements in the case of the Gaussian PSF are given in Table~\ref{tab}. Here, we have used a Taylor series expansion of \eq~\eqref{eq:cce-sliver} in $d$ for SLIVER, and used \eq~\eqref{eq:ce-di} for estimating the Chernoff exponent for direct imaging. 
The characteristic scalings with respect to $d$ of the conditional quantum Chernoff exponent and that of the three measurement schemes are shown in Fig.~\ref{fig:chernoff}. 
We  see that the Chernoff exponent of SLIVER agrees with the quantum limit for all practical purposes in the sub-Rayleigh regime $d \leq 1$.
\\

\noindent\textbf{Decision rule.}
In order to choose a hypothesis based on a sequence of B-SPADE/SLIVER observations, we need to fix a decision rule.
If the separation $d$ is known, the optimal decision rule is given by the likelihood-ratio test~\cite{VanTrees2013}:
For a given observation record  $(z_1,z_2,\ldots,z_M)$, we choose $\hp_2$ if $\prod_{j=1}^M \Lambda_2(z_j)/\Lambda_1(z_j) > p_1/p_2$, where $p_1$ and $p_2$ are the prior probabilities of $\hp_1$ and $\hp_2$ respectively, and choose $\hp_1$ otherwise.
If the separation is unknown, one can use the generalized-likelihood-ratio test~\cite{Kay1998}, which first estimates the separation and then does the likelihood-ratio test with the estimated value. 

Here, we propose a simplified decision rule that does not require the separation to be known or estimated.
Observe that if the detector corresponding to the modes orthogonal to the first mode in the three-mode basis (see \eq~(\ref{eq:phi1}--\ref{eq:phi3}) in Methods) clicked for any sample, we can infer  with certainty (in either source model) that two point sources are present, i.e., $\hp_2$ is true.
The simplified decision rule is then given by accepting $\hp_1$ only if this detector does not click during the entire observation period.

From Table~\ref{tab:probability} in the Methods, under the simplified decision rule, the false-alarm probability for $M$ samples is clearly $\alpha=0$ for both the B-SPADE and SLIVER measurement. 
The miss error probability is the probability that the detector corresponding to $\hp_2$ does not click, i.e.,
\begin{equation}
	\beta=[\Lambda_2\tsf{(off, off)} + \Lambda_2\tsf{(on, off)}]^M,
\end{equation}
where $\Lambda_2(\cdot,\cdot)$ is the probability of the measurement outcome under $\hp_2$.
It then can shown that $\beta^{\rm (meas)}=\exp(-M\xi^{\rm (meas)})$ for both the B-SPADE and SLIVER measurement.
\\

\noindent\textbf{Discussion.}
We have examined the problem of discriminating one thermal source from two closely separated ones for a given diffraction-limited imaging system. 
Using the exact thermal state of the image-plane field, we have derived the quantum Chernoff exponent for the detection problem.
We also have used the weak-source model of the image-plane field, which is very accurate in the optical regime due to the low brightness of a thermal source in each temporal mode, to obtain simple expressions for the Chernoff exponent.
The per-sample B-SPADE measurement that separates light in the PSF mode from the rest of the field was shown to attain the quantum-optimal Chernoff exponent for all values of two-source separation. 
Remarkably, it does so without the need for prior knowledge of the value of $d$, joint measurement over multiple modes, or photon-number resolution in each mode.
These properties are not shared by the quantum-optimal measurement elucidated by Helstrom~\cite{Helstrom1973}, which is not a structured receiver. These advantages also adhere to the SLIVER measurement, which is near-quantum-optimal in the sub-Rayleigh regime. Moreover, the experimental design of SLIVER is independent of the particular (reflection-symmetric) PSF of the imaging system.

In fact, the simplified decision rules proposed here for B-SPADE and SLIVER do not require resolving the arrival time of the detected photon or photons. To wit, only a single on-off detector without temporal resolution placed in the output corresponding to the modes orthogonal to the PSF (for B-SPADE) or
 to the antisymmetric component (for SLIVER) is sufficient for achieving the error probability behavior derived here.   Hypothesis $\hp_2$ is accepted if and only if this detector clicks at any time during the observation period.  If we need to simultaneously know
the conditional error probability, then at least two photon-number-resolving detectors (or gated on-off detectors with sufficient temporal resolution) are required such that the total number of the photons arriving on the image plane can be obtained from the observation.

Some implementation imperfections that can result in suboptimal performance deserve to be mentioned here.
We have so far assumed in our analysis that the optical axes of the B-SPADE and SLIVER devices are perfectly aligned to the source centroids.
In practice, we can use a portion of light to estimate the centroid before aligning the B-SPADE/SLIVER devices~\cite{Tsang2016b,Nair2016}.
From quantum parameter estimation theory, it is known that direct imaging can be used to achieve a good estimate of the centroid, provided that a sufficient number of photons had been collected~\cite{Tsang2016b}.
Remarkably, as long as the optical axes are perfectly aligned, the Chernoff exponents of both B-SPADE and SLIVER as well as the quantum limit are insensitive to the relative brightnesses under the weak source model, meaning that the advantage of B-SPADE and SLIVER over direct imaging still holds when the two hypothetical sources do not have equal brightness components.
Another possible source of imperfection is dark counts in the  photodetectors; this may affect the performance the B-SPADE and SLIVER schemes, especially those using the simplified decision rules.
To improve the robustness against dark counts or extraneous background light, we may use feedback strategies, like those developed in the context of distinguishing between optical coherent states~\cite{Dolinar1973,Geremia2004}. These issues will be addressed in subsequent work.

Although sophisticated optical microscopy techniques can help resolve multiple sources  better than direct imaging~\cite{WS15}, the manipulation of the source emission that they require is impossible in astronomical imaging for which the dominant detection technique is direct imaging. Our proof that the linear-optics schemes proposed here can yield Chernoff exponents that are orders of magnitude larger than that of direct imaging, coupled with  the rapid recent experimental progress on similar schemes~\cite{Tang2016,Yang2016,Tham2017,Paur2016},  holds out great promise for applications in astronomy and molecular imaging analysis in the near future.

\vspace{6 mm}
\noindent\textbf{\large Methods}

\noindent{\bf Density operators.} 
To calculate the Chernoff exponent and error probabilities, we need to express the density operators $\rho_1$ and $\rho_2$ in an appropriate basis.
We focus on a single temporal mode $\chi(t)$ of the image-plane field satisfying $\int_0^T \abs{\chi(t)}^2 \diff t =1$ on the observation interval $[0,T]$. 
The two mutually incoherent sources at $\pm \bd/2$ under $\tsf{H}_2$ are described by statistically independent zero-mean circular complex Gaussian amplitudes $A_1$ and $A_2$ with the probability density
\begin{align}
	\Pr(A_1,A_2) = \prod_{j=1}^2 \frac{1}{\pi\epsilon_j} \exp \lp -\abs{A_j}^2/\epsilon_j\rp,
\end{align}
where $\epsilon_1$ and $\epsilon_2$ are the brightnesses of the two point sources.
The image-plane field conditioned on $(A_1,A_2)$ is described quantum-mechanically as a coherent-state, i.e., an eigenstate $\ket{\phi_{A_1,A_2}}$ of the positive-frequency field operator $\hat{E}^{(+)}(\br,t)$ in the image plane: 
$\hat{E}^{(+)}(\br,t)\ket{\phi_{A_1,A_2}} = \mathcal{E}_{A_1,A_2}(\br)\chi(t) \ket{\phi_{A_1,A_2}}$, where $\mathcal{E}_{A_1,A_2}(\br)$ is given by
\begin{align} \label{eq:eigenfunc2}
	\mathcal{E}_{A_1,A_2}(\br) = A_1\psi\lp \br - \bd/2\rp + A_2\psi\lp\br + \bd/2\rp
\end{align}
and $\psi(\br)$ is the normalized PSF.
The density operator under $\hp_2$ is formally given by
\begin{align}
\rho_2 &= \int_{\mathbb{C}^2}\diff^2 A_1 \diff^2 A_2 \;\Pr(A_1,A_2) \ket{\phi_{A_1,A_2}}\bra{\phi_{A_1,A_2}}. 
\end{align}

Note that $\rho_2$ depends on the separation $d$ and is reduced to $\rho_1$ when setting $d=0$.
Moreover, it can be seen from \eq~\eqref{eq:eigenfunc2} that under $\hp_1$ we have $\mathcal{E}_{A_1,A_2}(\br) = (A_1+A_2)\psi(\br)$.
Thus, it is evident that, while the relevant coherent states are defined on a complete set of transverse-spatial modes on $\cl{I}$,   only three orthonormal modes are in excited (non-vacuum) states. 
These may be chosen as
\begin{align} 
\phi_1(\br) &= \psi(\br), \label{eq:phi1} \\
\phi_2(\br) &=  
\frac1{\sqrt{1-\mu^2}}
\lb \frac{\psi(\br -\bd/2) + \psi(\br + \bd/2)}{2\sqrt{\lambda_+}} - \mu\,\psi(\br)\rb, \label{eq:phi2}\\
\phi_3(\br) & = \frac{\psi(\br -\bd/2) - \psi(\br + \bd/2)}{2\sqrt{\lambda_-}}, \label{eq:phi3}
\end{align}
where $\lambda_\pm\equiv(1\pm\delta(d))/2$ and $\mu$ is given by \eq~\eqref{eq:lambda_mu}. 
Using $\{\phi_1(\br), \phi_2(\br), \phi_3(\br)\}$ as a spatial-mode basis, we have
\begin{equation}
	\mathcal{E}_{A_1,A_2}(\br) =  A_+\lb\mu\phi_1(\br) +\sqrt{1-\mu^2}\phi_2(\br)\rb + A_- \phi_3(\br),
\end{equation}
where $A_+ = \sqrt{\lambda_+}(A_1+A_2)$ and $A_- = \sqrt{\lambda_-}(A_1-A_2)$ are two circular complex Gaussian amplitudes random variables.
When $\epsilon_1=\epsilon_2=\epsilon/2$, the random variables $A_+$ and $A_-$ are statistically independent~\cite{Nair2016}.
In such a case, we get
\begin{align}
	\rho_1 &= \rho_{\tsf{th}}\lp  \epsilon\rp \otimes \ket{0}\bra{0} \otimes \ket{0}\bra{0},  
	\label{eq:rho1}\\
	\rho_2 &= U \lb \rho_{\tsf{th}}(\epsilon_+) \otimes \ket{0}\bra{0}\rb U^{\dag} \otimes \rho_{\tsf{th}}(\epsilon_-), 
	\label{eq:rho2}
\end{align} 
where
$\rho_{\tsf{th}}(\epsilon)=\sum_n [\epsilon^n / (\epsilon +1)^{n+1}]\ket n \bra n$
is the single-mode thermal state with $\epsilon$ average photons~\cite{MW95,Shapiro2009}, $U$ is a unitary beamsplitter transformation with transmissivity $\mu$ acting on the first two modes.
The $d$-dependent quantities $\epsilon_\pm$ and $\mu$ are given by \eq~\eqref{eq:lambda_mu}.
The transmissivity $\mu$ takes values in the range [-1,1] and equals unit when $d=0$. 
The beamsplitter implements the transformation
\begin{equation}\label{eq:bs}
	U\ket\alpha\otimes\ket\beta = |\mu \alpha - \sqrt{1-\mu^2}\beta \rangle 
						  \otimes |\mu \beta  + \sqrt{1-\mu^2}\alpha\rangle
\end{equation}
for input coherent states $\ket\alpha$ and $\ket\beta$, while for a number state-vacuum input $\ket{n}\ket{0}$, we have
\begin{equation}\label{eq:bsnumvac}
	U\ket n\otimes\ket 0 = \sum_{k=0}^n \sqrt{\binom{n}{k}} \,\mu^k \,(1 - \mu^2)^{\frac{n-k}{2}} \ket{k,n-k}.
\end{equation}
\\

\noindent\textbf{Quantum Chernoff exponent.} 
The quantum Chernoff exponent is given by $\xi = -\log\min_{0\leq s\leq1} Q_s$ with $Q_s\equiv\tr(\rho_1^s\rho_2^{1-s})$.
Using \eqs~\eqref{eq:rho1}, \eqref{eq:rho2}, and \eqref{eq:bsnumvac}, we get after some algebra:
\begin{align}
Q_s &=  \frac{1}{a p^s (1- b q^s)}, \label{Qs}
\end{align}
where the coefficients are 
\begin{align}\label{eq:coefficients}
\begin{split}
a &= (1+\epsilon_+)(1+\epsilon_-),\quad
b  = \frac{\epsilon_+}{1+\epsilon_+} \mu^2,\\
p &= \frac{1+\epsilon}{(1+\epsilon_+)(1+\epsilon_-)},\quad
q  = \frac{\epsilon(1+\epsilon_+)}{\epsilon_+(1+\epsilon)}.
\end{split}
\end{align}
It follows from $0\leq\epsilon_\pm\leq\epsilon$ and $\epsilon_++\epsilon_-=\epsilon$ that $p\leq 1$ and $q \geq 1$. Thus, $Q_s$ takes its minimum at $s=0$, i.e.,
\begin{align} 
	\min_{0\leq s \leq 1}Q_s &= Q_0 = \frac{1}{(1+\epsilon_-)(1+\epsilon_+ - \mu^2\epsilon_+)}, \label{QCB}
\end{align}
from which we obtain the quantum Chernoff exponent in \eq~\eqref{eq:ce-limit-bspade}.
\\

\begin{figure}[tb]
	\includegraphics{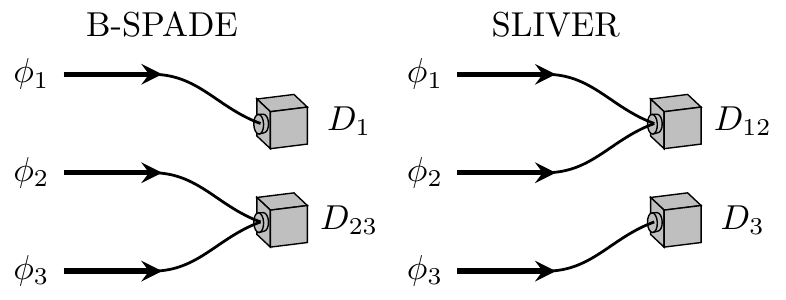}
	\caption{Effective description of the B-SPADE and SLIVER measurements in the three-mode basis. Here, $\phi_1$, $\phi_2$ and $\phi_3$ are the three orthonormal modes of \eqs~\eqref{eq:phi1}-\eqref{eq:phi3} used to represent the density operators in \eq~\eqref{eq:rho1} and \eqref{eq:rho2}.}
	\label{fig:detection}
\end{figure}

\begin{table*}[tb]
	\caption{Probability distributions of single-sample outcomes for the two on-off detectors used in the B-SPADE and SLIVER measurement. The first entry in each outcome refers to detection of the filtered PSF mode for B-SPADE and the symmetric field for SLIVER. The probability represented by $*$ can be obtained by the normalization of the probability distribution.
	}
	\label{tab:probability}
	\centering
	\begin{tabular}{l|c|cccc|}
	\hline

	\hline
	\multirow{2}*{Measurement} & \multirow{2}*{Hypothesis} &\multicolumn{4}{c|}{Probability of Event}\\
	\cline{3-6}
	&& \bigstrut \tsf{(off, off)} & \tsf{(on, off)} & \tsf{(off, on)} & \tsf{(on, on)}\\
	\hline
	\bigstrut \multirow{2}*{B-SPADE}  & $\hp_1$ &\bigstrut
		$\frac{1}{1+\epsilon}$ & $\frac{\epsilon}{1+\epsilon}$ & 0 & 0 \\
	& $\hp_2$ &\bigstrut 
		$\frac1{(1+\epsilon_+)(1+\epsilon_-)}$ &
		$\frac{\mu^2\epsilon_+}{(1+\epsilon_+)(1+\epsilon_-)(1+\epsilon_+-\mu^2\epsilon_+)}$ & 
		$\frac{1}{1+\mu^2\epsilon_+} - \frac{1}{(1+\epsilon_+)(1+\epsilon_-)}$ & $*$\\ 
	\hline
	\multirow{2}{1in}{SLIVER}  & $\hp_1$ &\bigstrut
		$\frac{1}{1+\epsilon}$ & $\frac{\epsilon}{1+\epsilon}$ & 0 & 0 \\
	& $\hp_2$ &\bigstrut
	 	$\frac1{(1+\epsilon_+)(1+\epsilon_-)}$ & $\frac{\epsilon_+}{(1+\epsilon_+)(1+\epsilon_-)}$ & 
	 	$\frac{\epsilon_-}{(1+\epsilon_+)(1+\epsilon_-)}$ & $\frac{\epsilon_+\epsilon_-}{(1+\epsilon_+)(1+\epsilon_-)}$\\ 
	\hline
	\end{tabular}
\end{table*}

\noindent\textbf{B-SPADE and SLIVER.} 
To calculate the Chernoff exponents as well as the error probabilities for  B-SPADE and SLIVER, it will be convenient to only focus on the effective action of the measurement on the relevant Hilbert subspace.
Figure~\ref{fig:detection} illustrates the effective actions of  B-SPADE and SLIVER on the mode subspace spanned by $\phi_1(\br)$, $\phi_2(\br)$ and $\phi_3(\br)$.
The B-SPADE measurement discriminates the first mode from the other two, while the SLIVER measurement discriminates the first two modes from the third, which is the sole excited antisymmetric mode.
We emphasize that our schemes are very different from a detector that resolves each of the modes $\phi_1(\br)$, $\phi_2(\br)$ and $\phi_3(\br)$, since implementing such a detector would require knowledge of $d$. Our schemes, on the other hand, work for any $d$.
Using \eqs~\eqref{eq:rho1} and \eqref{eq:rho2} and the effective action of the measurements shown in Fig.~\ref{fig:detection}, the probability distribution of measurement outcomes can be easily obtained as given in Table~\ref{tab:probability}, on which the calculation of the Chernoff exponents is based.

For  B-SPADE, we get
\begin{align}
	Q_s^{\rm (B-SPADE)}  
	= \frac{1+\tilde b \tilde q^s}{ap^s},
\end{align}
where $a$ and $p$ are given in \eq~\eqref{eq:coefficients}, $\tilde b=\mu^2\epsilon_+/(1+\epsilon_+-\mu^2\epsilon_+)$, and $\tilde q=\epsilon(1+\epsilon_+-\mu^2\epsilon_+)/\mu^2\epsilon_+\geq1$.
It follows that $Q_s$ is minimized over $[0,1]$ by taking $s=0$, leading to the result of $\xi^{\rm (B-SPADE)}$ in \eq~\eqref{eq:ce-limit-bspade}.
In the weak-source model, the probability distribution of a detected photon being at the two output ports is $\{1,0\}$ under $\hp_1$ and $\{\delta(d/2)^2,1-\delta(d/2)^2\}$ under $\hp_2$. Thus, one can easily obtain the result of $\xi_{\rm c}^{\rm (B-SPADE)}$ as shown in \eq~\eqref{eq:cce-bspade}.

For  SLIVER, the structure of $\rho_1$ and $\rho_2$ implies that the two detectors fire independently under both hypotheses. 
From these equations, the Chernoff exponent of SLIVER can be calculated (and corresponds to $s=0$ as for B-SPADE) with the result of \eq~\eqref{eq:ce-sliver}.
In the weak source model, the probability distribution of a detected photon being at the two output ports is $\{1,0\}$ under $\hp_1$ and $\{[\delta(d)+1]/2,[1-\delta(d)]/2\}$ under $\hp_2$. Thus, one can easily obtain the result of $\xi_{\rm c}^{\rm (SLIVER)}$ as shown in \eq~\eqref{eq:cce-sliver}.
\\

\noindent\textbf{Leading term of Chernoff exponent.} 
For a given measurement scheme, let $\Upsilon(z;d)$ be the resulting probability density of a measurement outcome $z$, where $d$ is the distance between the two hypothetic point sources.
The Chernoff exponent of \eq~\eqref{eq:classical_Chernoff} for testing  $\hp_2$ ($d>0$) against $\hp_1$ ($d=0$) can be written as $\xi(d) =-\log\min_{0\leq s\leq1}Q_s(d)$ with $Q_s(d) \equiv \int\mathrm{d}\mu(z)\Upsilon(z;0)^{1-s}\Upsilon(z;d)^s$.
Let us now focus on the leading term of $\xi(d)$ for small separations $d \approx 0$, where the optimal measurement performs much better than direct imaging.
We expand $Q_s(d)$ in a Taylor series as 
\begin{align}
	Q_s(d) &= 1 + \sum_{k=1}^\infty \frac{d^k}{k!} Q_{s,k}, \\
	Q_{s,k} &\equiv \int_\Omega\diff\mu(z)\, \Upsilon(z;0)^{1-s} 
	\left.\frac{\partial^k \Upsilon(z;d)^s}{\partial d^k} \right|_{d=0},
\end{align}
where $\Omega\equiv\{z \mid \Upsilon(z;0)>0\}$.
These coefficients $Q_{s,k}$ are independent of $d$.
Although in our model the separation $d$ is nonnegative, the PSFs in \eq~\eqref{eq:psf} can be easily extended to real numbers and meanwhile assured to be smooth at $d=0$.  
It then follows from \eq~\eqref{eq:eta} that $\Upsilon(z;d)=\Upsilon(z;-d)$ and thus all odd derivatives of $\Upsilon(z;d)$ with respect to $d$ at $d=0$ vanish for an arbitrary $z$. 
As a result, we have 
\begin{equation}
\begin{split}
	Q_{s,1} &= Q_{s,3}=0, \quad Q_{s,2}=sg^{(2)}(0), \\
	Q_{s,4} &= 3s(s-1)\mathcal{K} + s g^{(4)}(0),
\end{split}
\end{equation}
where $g^{(k)}$ denotes the $k$-th derivatives of $g(d)\equiv\int_\Omega\diff\mu(z)\,\Upsilon(z;d)$ and 
\begin{equation}
\mathcal K\equiv\int_\Omega\diff\mu(z)\,\frac1{\Upsilon(z;0)}  
\left[\frac{\partial^2\Upsilon(z;d)}{\partial d^2}\right]_{d=0}^2.
\end{equation}

For the ideal direct imaging scenario in the weak-source model, the measurement outcome $z$ is the coordinates of a detected photon in the image plane, i.e., $z=(x,y) \in \cl{I}$.
Suppose that the two point sources are aligned along the $x$-axis, $\Upsilon(z;d)$ is then given by \eq~\eqref{eq:di} with $z=(x,y)$.
For all three typical kinds of PSFs considered in this work, we have $g(d)=1$ for direct imaging.
In such a case, the leading $d$-dependent term in the Taylor series of $Q_s(d)$ is of fourth order. 
It then follows that $\xi_{\rm c}^{\rm (DI)} \simeq \frac{d^4}{32} \mathcal{K}$, which is \eq~\eqref{eq:ce-di}.
On the other hand, for B-SPADE and SLIVER, it can be seen from Table~\ref{tab:probability} that $g(d)=\Lambda_2(\tsf{off,off})+\Lambda_2(\tsf{on,off})$, so that $g^{(2)}(0)$ is nonzero and the leading term in $Q_{s}(d)$ is second-order in $d$.
\\

\bibliographystyle{naturemag}
\bibliography{onevstwo_joint}

\begin{thebibliography}{10}
\expandafter\ifx\csname url\endcsname\relax
  \def\url#1{\texttt{#1}}\fi
\expandafter\ifx\csname urlprefix\endcsname\relax\def\urlprefix{URL }\fi
\providecommand{\bibinfo}[2]{#2}
\providecommand{\eprint}[2][]{\url{#2}}

\bibitem{LordRayleigh1879}
\bibinfo{author}{Lord~Rayleigh, F. R.~S.}
\newblock \bibinfo{title}{Xxxi. investigations in optics, with special
  reference to the spectroscope}.
\newblock \emph{\bibinfo{journal}{Philosophical Magazine Series 5}}
  \textbf{\bibinfo{volume}{8}}, \bibinfo{pages}{261--274}
  (\bibinfo{year}{1879}).

\bibitem{Feynman1963}
\bibinfo{author}{Feynman, R.}, \bibinfo{author}{Leighton, R.} \&
  \bibinfo{author}{Sands, M.}
\newblock \emph{\bibinfo{title}{The Feynman Lectures on Physics:Volume I}}
  (\bibinfo{publisher}{Addison-Wesley}, \bibinfo{year}{1963}).

\bibitem{Ram2006}
\bibinfo{author}{Ram, S.}, \bibinfo{author}{Ward, E.~S.} \&
  \bibinfo{author}{Ober, R.~J.}
\newblock \bibinfo{title}{Beyond {Rayleigh's} criterion: A resolution measure
  with application to single-molecule microscopy}.
\newblock \emph{\bibinfo{journal}{Proc. Natl. Acad. Sci. U.S.A.}}
  \textbf{\bibinfo{volume}{103}}, \bibinfo{pages}{4457--4462}
  (\bibinfo{year}{2006}).

\bibitem{Chao2016}
\bibinfo{author}{Chao, J.}, \bibinfo{author}{Ward, E.~S.} \&
  \bibinfo{author}{Ober, R.~J.}
\newblock \bibinfo{title}{Fisher information theory for parameter estimation in
  single molecule microscopy: tutorial}.
\newblock \emph{\bibinfo{journal}{J. Opt. Soc. Am. A}}
  \textbf{\bibinfo{volume}{33}}, \bibinfo{pages}{B36--B57}
  (\bibinfo{year}{2016}).

\bibitem{Helstrom1976}
\bibinfo{author}{Helstrom, C.~W.}
\newblock \emph{\bibinfo{title}{Quantum Detection and Estimation Theory}}
  (\bibinfo{publisher}{Academic Press, New York}, \bibinfo{year}{1976}).

\bibitem{Holevo1982}
\bibinfo{author}{Holevo, A.~S.}
\newblock \emph{\bibinfo{title}{Probabilistic and Statistical Aspects of
  Quantum Theory}} (\bibinfo{publisher}{North-Holland, Amsterdam},
  \bibinfo{year}{1982}).

\bibitem{Tsang2016b}
\bibinfo{author}{Tsang, M.}, \bibinfo{author}{Nair, R.} \& \bibinfo{author}{Lu,
  X.-M.}
\newblock \bibinfo{title}{Quantum theory of superresolution for two incoherent
  optical point sources}.
\newblock \emph{\bibinfo{journal}{Phys. Rev. X}} \textbf{\bibinfo{volume}{6}},
  \bibinfo{pages}{031033} (\bibinfo{year}{2016}).

\bibitem{Nair2016}
\bibinfo{author}{Nair, R.} \& \bibinfo{author}{Tsang, M.}
\newblock \bibinfo{title}{Interferometric superlocalization of two incoherent
  optical point sources}.
\newblock \emph{\bibinfo{journal}{Opt. Express}} \textbf{\bibinfo{volume}{24}},
  \bibinfo{pages}{3684--3701} (\bibinfo{year}{2016}).

\bibitem{Tsang2016a}
\bibinfo{author}{Tsang, M.}, \bibinfo{author}{Nair, R.} \& \bibinfo{author}{Lu,
  X.-M.}
\newblock \bibinfo{title}{Semiclassical theory of superresolution for two
  incoherent optical point sources}  (\bibinfo{year}{2016}).
\newblock \eprint{1602.04655}.

\bibitem{Nair2016a}
\bibinfo{author}{Nair, R.} \& \bibinfo{author}{Tsang, M.}
\newblock \bibinfo{title}{Far-field superresolution of thermal electromagnetic
  sources at the quantum limit}.
\newblock \emph{\bibinfo{journal}{Phys. Rev. Lett.}}
  \textbf{\bibinfo{volume}{117}}, \bibinfo{pages}{190801}
  (\bibinfo{year}{2016}).

\bibitem{Ang2016}
\bibinfo{author}{Ang, S.~Z.}, \bibinfo{author}{Nair, R.} \&
  \bibinfo{author}{Tsang, M.}
\newblock \bibinfo{title}{Quantum limit for two-dimensional resolution of two
  incoherent optical point sources}.
\newblock \emph{\bibinfo{journal}{Phys. Rev. A}} \textbf{\bibinfo{volume}{95}},
  \bibinfo{pages}{063847} (\bibinfo{year}{2017}).

\bibitem{Lupo2016}
\bibinfo{author}{Lupo, C.} \& \bibinfo{author}{Pirandola, S.}
\newblock \bibinfo{title}{Ultimate precision bound of quantum and subwavelength
  imaging}.
\newblock \emph{\bibinfo{journal}{Phys. Rev. Lett.}}
  \textbf{\bibinfo{volume}{117}}, \bibinfo{pages}{190802}
  (\bibinfo{year}{2016}).

\bibitem{Rehacek2016}
\bibinfo{author}{Rehacek, J.} \emph{et~al.}
\newblock \bibinfo{title}{Dispelling {R}ayleigh's curse} \eprint{1607.05837}.

\bibitem{Tsang2016c}
\bibinfo{author}{Tsang, M.}
\newblock \bibinfo{title}{Subdiffraction incoherent optical imaging via
  spatial-mode demultiplexing}.
\newblock \emph{\bibinfo{journal}{New Journal of Physics}}
  \textbf{\bibinfo{volume}{19}}, \bibinfo{pages}{023054}
  (\bibinfo{year}{2017}).

\bibitem{KGA17}
\bibinfo{author}{Kerviche, R.}, \bibinfo{author}{Guha, S.} \&
  \bibinfo{author}{Ashok, A.}
\newblock \bibinfo{title}{Fundamental limit of resolving two point sources
  limited by an arbitrary point spread function}.
\newblock In \emph{\bibinfo{booktitle}{2017 IEEE International Symposium on
  Information Theory (ISIT)}}, \bibinfo{pages}{441--445}
  (\bibinfo{year}{2017}).

\bibitem{YNT+17}
\bibinfo{author}{Yang, F.}, \bibinfo{author}{Nair, R.}, \bibinfo{author}{Tsang,
  M.}, \bibinfo{author}{Simon, C.} \& \bibinfo{author}{Lvovsky, A.~I.}
\newblock \bibinfo{title}{Fisher information for far-field linear optical
  superresolution via homodyne or heterodyne detection in a higher-order local
  oscillator mode}.
\newblock \emph{\bibinfo{journal}{Phys. Rev. A}} \textbf{\bibinfo{volume}{96}},
  \bibinfo{pages}{063829} (\bibinfo{year}{2017}).

\bibitem{Tang2016}
\bibinfo{author}{Tang, Z.~S.}, \bibinfo{author}{Durak, K.} \&
  \bibinfo{author}{Ling, A.}
\newblock \bibinfo{title}{Fault-tolerant and finite-error localization for
  point emitters within the diffraction limit}.
\newblock \emph{\bibinfo{journal}{Opt. Express}} \textbf{\bibinfo{volume}{24}},
  \bibinfo{pages}{22004--22012} (\bibinfo{year}{2016}).

\bibitem{Yang2016}
\bibinfo{author}{Yang, F.}, \bibinfo{author}{Nair, R.}, \bibinfo{author}{Tsang,
  M.}, \bibinfo{author}{Simon, C.} \& \bibinfo{author}{Lvovsky, A.~I.}
\newblock \bibinfo{title}{Fisher information for far-field linear optical
  superresolution via homodyne or heterodyne detection in a higher-order local
  oscillator mode}.
\newblock \emph{\bibinfo{journal}{Phys. Rev. A}} \textbf{\bibinfo{volume}{96}},
  \bibinfo{pages}{063829} (\bibinfo{year}{2017}).

\bibitem{Tham2017}
\bibinfo{author}{Tham, W.-K.}, \bibinfo{author}{Ferretti, H.} \&
  \bibinfo{author}{Steinberg, A.~M.}
\newblock \bibinfo{title}{Beating {R}ayleigh's curse by imaging using phase
  information}.
\newblock \emph{\bibinfo{journal}{Phys. Rev. Lett.}}
  \textbf{\bibinfo{volume}{118}}, \bibinfo{pages}{070801}
  (\bibinfo{year}{2017}).

\bibitem{Paur2016}
\bibinfo{author}{Pa\'{u}r, M.}, \bibinfo{author}{Stoklasa, B.},
  \bibinfo{author}{Hradil, Z.}, \bibinfo{author}{S\'{a}nchez-Soto, L.~L.} \&
  \bibinfo{author}{Rehacek, J.}
\newblock \bibinfo{title}{Achieving the ultimate optical resolution}.
\newblock \emph{\bibinfo{journal}{Optica}} \textbf{\bibinfo{volume}{3}},
  \bibinfo{pages}{1144--1147} (\bibinfo{year}{2016}).

\bibitem{Harris1964}
\bibinfo{author}{Harris, J.~L.}
\newblock \bibinfo{title}{Resolving power and decision theory}.
\newblock \emph{\bibinfo{journal}{J. Opt. Soc. Am.}}
  \textbf{\bibinfo{volume}{54}}, \bibinfo{pages}{606--611}
  (\bibinfo{year}{1964}).

\bibitem{Helstrom1973}
\bibinfo{author}{Helstrom, C.}
\newblock \bibinfo{title}{Resolution of point sources of light as analyzed by
  quantum detection theory}.
\newblock \emph{\bibinfo{journal}{IEEE Trans. Inform. Theory}}
  \textbf{\bibinfo{volume}{19}}, \bibinfo{pages}{389--398}
  (\bibinfo{year}{1973}).

\bibitem{Acuna1997}
\bibinfo{author}{Acuna, C.~O.} \& \bibinfo{author}{Horowitz, J.}
\newblock \bibinfo{title}{A statistical approach to the resolution of point
  sources}.
\newblock \emph{\bibinfo{journal}{J. Appl. Statist.}}
  \textbf{\bibinfo{volume}{24}}, \bibinfo{pages}{421--436}
  (\bibinfo{year}{1997}).

\bibitem{Shahram2006}
\bibinfo{author}{Shahram, M.} \& \bibinfo{author}{Milanfar, P.}
\newblock \bibinfo{title}{Statistical and information-theoretic analysis of
  resolution in imaging}.
\newblock \emph{\bibinfo{journal}{IEEE Trans. Inform. Theor.}}
  \textbf{\bibinfo{volume}{52}}, \bibinfo{pages}{3411--3437}
  (\bibinfo{year}{2006}).

\bibitem{Dutton2010}
\bibinfo{author}{Dutton, Z.}, \bibinfo{author}{Shapiro, J.~H.} \&
  \bibinfo{author}{Guha, S.}
\newblock \bibinfo{title}{Ladar resolution improvement using receivers enhanced
  with squeezed-vacuum injection and phase-sensitive amplification}.
\newblock \emph{\bibinfo{journal}{J. Opt. Soc. Am. B}}
  \textbf{\bibinfo{volume}{27}}, \bibinfo{pages}{A63--A72}
  (\bibinfo{year}{2010}).

\bibitem{Labeyrie2006}
\bibinfo{author}{Labeyrie, A.}, \bibinfo{author}{Lipson, S.~G.} \&
  \bibinfo{author}{Nisenson, P.}
\newblock \emph{\bibinfo{title}{An Introduction to Optical Stellar
  Interferometry}} (\bibinfo{publisher}{Cambridge University Press},
  \bibinfo{year}{2006}).

\bibitem{Nan2013}
\bibinfo{author}{Nan, X.} \emph{et~al.}
\newblock \bibinfo{title}{Single-molecule superresolution imaging allows
  quantitative analysis of raf multimer formation and signaling}.
\newblock \emph{\bibinfo{journal}{Proc. Natl. Acad. Sci. U.S.A.}}
  \textbf{\bibinfo{volume}{110}}, \bibinfo{pages}{18519--18524}
  (\bibinfo{year}{2013}).

\bibitem{Hayashi2006}
\bibinfo{author}{Hayashi, M.}
\newblock \emph{\bibinfo{title}{Quantum Information: An Introduction}}
  (\bibinfo{publisher}{Springer-Verlag, Berlin Heidelberg},
  \bibinfo{year}{2006}), \bibinfo{edition}{1} edn.

\bibitem{Holevo1973b}
\bibinfo{author}{Holevo, A.}
\newblock \bibinfo{title}{Statistical decision theory for quantum systems}.
\newblock \emph{\bibinfo{journal}{J. Multivar. Anal.}}
  \textbf{\bibinfo{volume}{3}}, \bibinfo{pages}{337 -- 394}
  (\bibinfo{year}{1973}).

\bibitem{RZB+94}
\bibinfo{author}{Reck, M.}, \bibinfo{author}{Zeilinger, A.},
  \bibinfo{author}{Bernstein, H.~J.} \& \bibinfo{author}{Bertani, P.}
\newblock \bibinfo{title}{Experimental realization of any discrete unitary
  operator}.
\newblock \emph{\bibinfo{journal}{Phys. Rev. Lett.}}
  \textbf{\bibinfo{volume}{73}}, \bibinfo{pages}{58--61}
  (\bibinfo{year}{1994}).

\bibitem{MNJ+10}
\bibinfo{author}{Morizur, J.-F.} \emph{et~al.}
\newblock \bibinfo{title}{Programmable unitary spatial mode manipulation}.
\newblock \emph{\bibinfo{journal}{J. Opt. Soc. Am. A}}
  \textbf{\bibinfo{volume}{27}}, \bibinfo{pages}{2524--2531}
  (\bibinfo{year}{2010}).

\bibitem{Mil13}
\bibinfo{author}{Miller, D. A.~B.}
\newblock \bibinfo{title}{Reconfigurable add-drop multiplexer for spatial
  modes}.
\newblock \emph{\bibinfo{journal}{Opt. Express}} \textbf{\bibinfo{volume}{21}},
  \bibinfo{pages}{20220--20229} (\bibinfo{year}{2013}).

\bibitem{chernoff1952}
\bibinfo{author}{Chernoff, H.}
\newblock \bibinfo{title}{A measure of asymptotic efficiency for tests of a
  hypothesis based on the sum of observations}.
\newblock \emph{\bibinfo{journal}{Ann. Math. Statist.}}
  \textbf{\bibinfo{volume}{23}}, \bibinfo{pages}{493--507}
  (\bibinfo{year}{1952}).

\bibitem{VanTrees2013}
\bibinfo{author}{Van~Trees, H.~L.}, \bibinfo{author}{Bell, K.~L.} \&
  \bibinfo{author}{Tian, Z.}
\newblock \emph{\bibinfo{title}{Detection, Estimation, and Modulation Theory,
  Part I}} (\bibinfo{publisher}{Wiley}, \bibinfo{year}{2013}),
  \bibinfo{edition}{2nd edition} edn.

\bibitem{Cover2012}
\bibinfo{author}{Cover, T.} \& \bibinfo{author}{Thomas, J.}
\newblock \emph{\bibinfo{title}{Elements of Information Theory}}
  (\bibinfo{publisher}{Wiley}, \bibinfo{year}{2006}), \bibinfo{edition}{2nd
  edition} edn.

\bibitem{Ogawa2004}
\bibinfo{author}{Ogawa, T.} \& \bibinfo{author}{Hayashi, M.}
\newblock \bibinfo{title}{On error exponents in quantum hypothesis testing}.
\newblock \emph{\bibinfo{journal}{IEEE Transactions on Information Theory}}
  \textbf{\bibinfo{volume}{50}}, \bibinfo{pages}{1368--1372}
  (\bibinfo{year}{2004}).

\bibitem{Kargin2005}
\bibinfo{author}{Kargin, V.}
\newblock \bibinfo{title}{On the {Chernoff} bound for efficiency of quantum
  hypothesis testing}.
\newblock \emph{\bibinfo{journal}{Ann. Stat.}} \textbf{\bibinfo{volume}{33}},
  \bibinfo{pages}{959--976} (\bibinfo{year}{2005}).

\bibitem{Audenaert2007}
\bibinfo{author}{Audenaert, K. M.~R.} \emph{et~al.}
\newblock \bibinfo{title}{Discriminating states: The quantum {Chernoff} bound}.
\newblock \emph{\bibinfo{journal}{Phys. Rev. Lett.}}
  \textbf{\bibinfo{volume}{98}}, \bibinfo{pages}{160501}
  (\bibinfo{year}{2007}).

\bibitem{Nussbaum2009}
\bibinfo{author}{Nussbaum, M.} \& \bibinfo{author}{Szko\l{}a, A.}
\newblock \bibinfo{title}{The {Chernoff} lower bound for symmetric quantum
  hypothesis testing}.
\newblock \emph{\bibinfo{journal}{Ann. Statist.}}
  \textbf{\bibinfo{volume}{37}}, \bibinfo{pages}{1040--1057}
  (\bibinfo{year}{2009}).

\bibitem{Audenaert2008}
\bibinfo{author}{Audenaert, K.}, \bibinfo{author}{Nussbaum, M.},
  \bibinfo{author}{Szko\l{}a, A.} \& \bibinfo{author}{Verstraete, F.}
\newblock \bibinfo{title}{Asymptotic error rates in quantum hypothesis
  testing}.
\newblock \emph{\bibinfo{journal}{Commun. Math. Phys.}}
  \textbf{\bibinfo{volume}{279}}, \bibinfo{pages}{251--283}
  (\bibinfo{year}{2008}).

\bibitem{Goo05Fourier}
\bibinfo{author}{Goodman, J.~W.}
\newblock \emph{\bibinfo{title}{Introduction to Fourier Optics}}
  (\bibinfo{publisher}{Roberts and Company Publishers}, \bibinfo{year}{2005}),
  \bibinfo{edition}{3rd} edn.

\bibitem{Goo85Statistical}
\bibinfo{author}{Goodman, J.~W.}
\newblock \emph{\bibinfo{title}{Statistical Optics}} (\bibinfo{publisher}{John
  Wiley \& Sons}, \bibinfo{year}{1985}).

\bibitem{Gottesman2012}
\bibinfo{author}{Gottesman, D.}, \bibinfo{author}{Jennewein, T.} \&
  \bibinfo{author}{Croke, S.}
\newblock \bibinfo{title}{Longer-baseline telescopes using quantum repeaters}.
\newblock \emph{\bibinfo{journal}{Phys. Rev. Lett.}}
  \textbf{\bibinfo{volume}{109}}, \bibinfo{pages}{070503}
  (\bibinfo{year}{2012}).

\bibitem{Tsang2011}
\bibinfo{author}{Tsang, M.}
\newblock \bibinfo{title}{Quantum nonlocality in weak-thermal-light
  interferometry}.
\newblock \emph{\bibinfo{journal}{Phys. Rev. Lett.}}
  \textbf{\bibinfo{volume}{107}}, \bibinfo{pages}{270402}
  (\bibinfo{year}{2011}).

\bibitem{Kay1998}
\bibinfo{author}{Kay, S.~M.}
\newblock \emph{\bibinfo{title}{Fundamentals of Statistical Signal Processing,
  Volume II: Detection Theory}} (\bibinfo{publisher}{Prentice Hall, New
  Jersey}, \bibinfo{year}{1998}), \bibinfo{edition}{1st} edn.

\bibitem{Dolinar1973}
\bibinfo{author}{Dolinar, S.~J.}
\newblock \bibinfo{title}{An optimum receiver for the binary coherent state
  quantum channel}.
\newblock \emph{\bibinfo{journal}{MIT Res. Lab. Electron. Quart. Progr. Rep.}}
  \textbf{\bibinfo{volume}{111}}, \bibinfo{pages}{115--120}
  (\bibinfo{year}{1973}).

\bibitem{Geremia2004}
\bibinfo{author}{Geremia, J.}
\newblock \bibinfo{title}{Distinguishing between optical coherent states with
  imperfect detection}.
\newblock \emph{\bibinfo{journal}{Phys. Rev. A}} \textbf{\bibinfo{volume}{70}},
  \bibinfo{pages}{062303} (\bibinfo{year}{2004}).

\bibitem{WS15}
\bibinfo{author}{Weisenburger, S.} \& \bibinfo{author}{Sandoghdar, V.}
\newblock \bibinfo{title}{Light microscopy: an ongoing contemporary
  revolution}.
\newblock \emph{\bibinfo{journal}{Contemporary Physics}}
  \textbf{\bibinfo{volume}{56}}, \bibinfo{pages}{123--143}
  (\bibinfo{year}{2015}).

\bibitem{MW95}
\bibinfo{author}{Mandel, L.} \& \bibinfo{author}{Wolf, E.}
\newblock \emph{\bibinfo{title}{Optical Coherence and Quantum Optics}}
  (\bibinfo{publisher}{Cambridge University Press, Cambridge},
  \bibinfo{year}{1995}).

\bibitem{Shapiro2009}
\bibinfo{author}{Shapiro, J.~H.}
\newblock \bibinfo{title}{The quantum theory of optical communications}.
\newblock \emph{\bibinfo{journal}{IEEE Journal of Selected Topics in Quantum
  Electronics}} \textbf{\bibinfo{volume}{15}}, \bibinfo{pages}{1547 --1569}
  (\bibinfo{year}{2009}).

\bibitem{Krovi2016}
\bibinfo{author}{Krovi, H.}, \bibinfo{author}{Guha, S.} \&
  \bibinfo{author}{Shapiro, J.~H.}
\newblock \bibinfo{title}{Attaining the quantum limit of passive imaging}
  (\bibinfo{year}{2016}).
\newblock \eprint{arXiv:1609.00684}.

\bibitem{Lu2016}
\bibinfo{author}{Lu, X.-M.}, \bibinfo{author}{Nair, R.} \&
  \bibinfo{author}{Tsang, M.}
\newblock \bibinfo{title}{Quantum-optimal detection of one-versus-two
  incoherent sources with arbitrary separation}  (\bibinfo{year}{2016}).
\newblock \eprint{arXiv:1609.03025}.

\end{thebibliography}

\vspace{8pt}
\noindent\textbf{Acknowledgments}\\  
We thank Mankei Tsang for several useful discussions.
This work was supported by the Singapore National Research Foundation under NRF Grant No.~NRF-NRFF2011-07,
the Singapore Ministry of Education Academic Research Fund Tier 1 Project R-263-000-C06-112,
the Natural Science Foundation of Zhejiang Province of China under Grant No.~LY18A050003, 
the Defense Advanced Research Projects Agency's (DARPA) Information in a Photon (InPho) program under Contract No.~HR0011-10-C-0159, the REVEAL and EXTREME Imaging program, 
and the Air Force Office of Scientific Research under Grant No.~FA9550-14-1-0052.

\vspace{8pt}
\noindent\textbf{Author contributions}\\
X.-M.~L. performed the weak-source model analysis. 
H.~K., S.~G., J.~H.~S., and R.~N. performed the thermal-state analysis. 
R.~N. performed the SLIVER analyses. 
X.-M.~L. and R.~N. wrote the manuscript.
All the authors discussed extensively during the course of this work. This paper extends and unifies preliminary work  in the preprints~\cite{Krovi2016} (thermal-state model) and~\cite{Lu2016} (weak-source model).
\\

\noindent{\bf Competing financial interests}: 
The authors declare no competing financial interests.
\end{document}